\documentclass[usenatbib]{mn2e}
\usepackage{graphicx}
\usepackage{ifthen}
\usepackage{url}

\def\ltsima{$\; \buildrel < \over \sim \;$}
\def\lta{\lower.5ex\hbox{\ltsima}}
\def\gtsima{$\; \buildrel > \over \sim \;$}
\def\simgt{\lower.5ex\hbox{\gtsima}}
\def\kms{{\rm\,km\,s^{-1}}}

\def\msun{{\rm\,M_\odot}}
\def\lsun{{\rm\,L_\odot}}

\def\AA{$\; \buildrel \circ \over {\rm A}$}

\def\s{\ifmmode \widetilde \else \~\fi}
\def\={\overline}

\def\spose#1{\hbox to 0pt{#1\hss}}

\def\lta{\mathrel{\spose{\lower 3pt\hbox{$\mathchar"218$}}
     \raise 2.0pt\hbox{$\mathchar"13C$}}}
\def\gta{\mathrel{\spose{\lower 3pt\hbox{$\mathchar"218$}}
     \raise 2.0pt\hbox{$\mathchar"13E$}}}
\def\Dt{\spose{\raise 1.5ex\hbox{\hskip3pt$\mathchar"201$}}}    
\def\dt{\spose{\raise 1.0ex\hbox{\hskip2pt$\mathchar"201$}}}    

\def\dotsfill{\leaders\hbox to 1em{\hss.\hss}\hfill}

\def\FeH{{\rm[Fe/H]}}

\def\andxi{And\,XI}
\def\andxii{And\,XII}
\def\andxiii{And\,XIII}

\def\streamc{Stream\,`C'}
\def\streamcr{Stream\,`Cr'}
\def\streamcp{Stream\,`Cp'}
\def\streamd{Stream\,`D'}
\def\ec4{EC4}

\loadboldmathitalic 
\title[]{
A spectroscopic survey of EC4, an Extended Cluster in Andromeda's halo$^{1}$
}
\author[M.\ Collins et al.] 
{M.\ L.\ M.\ Collins$^2$, S.\ C.\ Chapman$^{2}$, M.\ Irwin$^{2}$, R.\ Ibata$^{3}$, N.\ F.\ Martin$^{4}$, 
\newauthor
 A.\ M.\ N.\ Ferguson$^{5}$, A.\ Huxor$^{9}$,  G.\ F.\ Lewis$^{6}$, A.\ D.\ Mackey$^5$, A.\ W.\ McConnachie$^{7}$, 
\newauthor
  N.\ Tanvir$^{8}$ \\
$^{2}$ Institute of Astronomy, Madingley Road, Cambridge, CB3 0HA, U.K.\\
$^{3}$ Observatoire de Strasbourg, 11, rue de l'Universit\'e, F-67000, Strasbourg, France\\
$^{4}$ Max-Planck-Institut f\"ur Astronomie, K\"onigstuhl 17, D-69117 
Heidelberg, Germany\\
$^{5}$ Institute for Astronomy, University of Edinburgh, Royal Observatory, Blackford Hill, Edinburgh, UK EH9~3HJ\\
$^{6}$ Sydney Institute for Astronomy, School of Physics, A29, University of Sydney, NSW 2006, Australia\\
$^{7}$ NRC Herzberg Institute of Astrophysics, 5071 West Saanich Road, Victoria, V9E 2E7, Canada\\
$^{8}$ Department of Physics \& Astronomy, University of Leicester, Leicester, LE17RH, UK\\ 
$^{9}$ Department of Physics, University of Bristol, Tyndall Ave, Bristol BS8 1TL\\
}

\date{\today}
\begin{document} 
\maketitle 

\begin{abstract} 
We present a spectroscopic survey of candidate red giant branch stars in the extended star cluster, \ec4, discovered in the halo of M31 from our CFHT/MegaCam survey, overlapping the tidal streams, \streamcp\ and \streamcr. 
These observations used the DEep Imaging Multi-Object Spectrograph (DEIMOS) mounted on
the Keck~II telescope to obtain spectra around the Ca{\sc II} triplet region with $\sim1.3$\AA\ resolution.
Six stars lying on the red giant branch within 2 core-radii of the centre of \ec4 are found to have an average 
$v_{r}=-287.9^{+1.9}_{-2.4}$~kms$^{-1}$ and $\sigma_{v, corr}=2.7^{+4.2}_{-2.7}$~kms$^{-1}$, taking instrumental errors into account. The resulting mass-to-light ratio for EC4 is M/L$=6.7^{+15}_{-6.7}\msun/\lsun$, a value that is consistent with a globular cluster within the $1\sigma$ errors we derive. From the summed spectra of our member stars, we find EC4 to be metal-poor, with [Fe/H]=-1.6$\pm0.15$. We discuss several formation and evolution scenarios which could account for our kinematic and metallicity constraints on EC4, and conclude that EC4 is most comparable with an extended globular cluster. We also compare the kinematics and metallicity of EC4 with \streamcp\ and \streamcr, and find that EC4 bears a striking resemblance to \streamcp\ in terms  of velocity, and that the two structures are identical in terms of both their spectroscopic and photometric metallicities. From this we conclude that EC4 is likely related to \streamcp.
\end{abstract}
 

\begin{figure*}
\begin{center}
\includegraphics[angle=0,width=0.395\hsize]{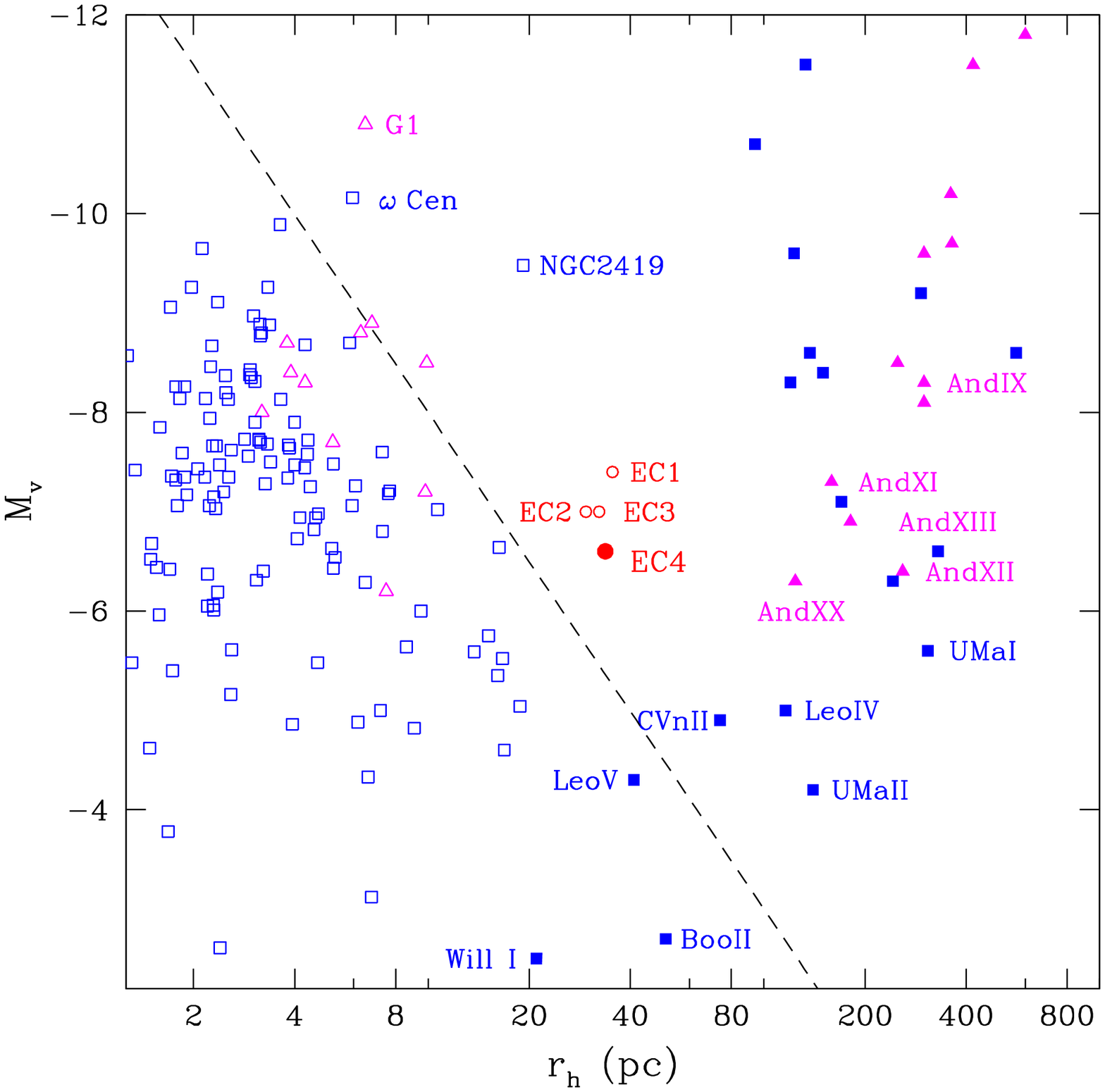}
\caption{
The half light radius and luminosity for Extended Clusters EC4 (red circle) and ECs 1-3(open red circles, \citet{mackey06}) compared with the faintest known M31 dwarf galaxies (filled pink triangles, \citet{mcconnachie06,mcconnachie05a,zucker04,irwin08,mcconnachie08}, and Collins et al. in prep.), M31 globular clusters (open pink triangles, \citet{mackey07}), MW associated dSphs (filled blue squares, \citet{irwin95,willman06,simon07,irwin07,belokurov07,belokurov08,martin08}) and MW GCs (open blue squares, \citet{vanden96}). In addition, the line of the equation $\log r_{h}=0.2M_{v}+2.6$ (from \citet{vanden04}) is plotted as a dashed line, above which only two Galactic GCs ($\omega$-Cen and NGC 2419) are found.
}
\label{lsize}
\end{center}
\end{figure*}

\section{Introduction}
\footnotetext[1]{The data presented herein were obtained at the W.M. Keck Observatory, which is operated as a scientific
partnership among the California Institute of Technology, the University of California and the National Aeronautics and
Space Administration. The Observatory was made possible by the generous financial support of the W.M. Keck Foundation.}

Over the past few years, the number, depth and coverage of both photometric and spectroscopic surveys of the Local Group has increased dramatically, resulting in the discovery of a myriad of globular clusters (GCs) and dwarf spheroidal galaxies (dSphs) orbiting within its gravitational potential. And whilst traditionally, satellite objects such as these have been somewhat simple to classify in terms of their half-light radii (r$_h\sim$2-5pc for GCs cf. $\sim$few 100pc for dSphs) and their dark matter content via the calculation of their mass-to-light (M/L) ratios (M/L$\lta3$ for GCs, which are not considered to contain dark matter, cf. M/L$\gta$10 for the dark matter dominated dSphs), recent discoveries of low-surface brightness objects with properties common to both GCs and dSphs have caused this historical divide to blur. SDSS has been instrumental in this process within the confines of our own Milky Way (MW), where in addition to a host of dSphs, a number of low luminosity objects have been discovered out in the halo with properties that see them encroaching on the void in the size-luminosity parameter space between GCs and dSphs \citep{willman05a,belokurov07,simon07,walsh07,belokurov08}. Similarly, wide-field photometric surveys of M31 and its extensive halo conducted with the Isaac Newton Telescope Wide-Field Camera and CFHT MegaCam \citep{ibata01b,ferguson02,ibata07} have revealed equally puzzling entities \citep{huxor05,mackey06,huxor08}, as well as vast numbers of more ``typical'' GCs and dSphs \citep{chapman05,martin06,ibata07,huxor08,irwin08,mcconnachie08}.

Some of the more notable of these recent MW discoveries include the unusual objects Willman I, Segue I, Bootes II and Leo V, all of which are low luminoisty objects (-2$>M_V>$-4.5), with half light radii between $\sim20-40$pc \citep{willman05a,belokurov07,walsh07,belokurov08}. There is also some debate as to whether such objects are dark matter dominated, which would suggest that they are members of the dSph population or if they are merely unusually extended clusters, with no sizeable dark matter component. In the case of Bootes II and Willman I, these unusual combinations of size and luminosity are also accompanied by internal metallicity dispersions and popualtion variations (though in the case of Willman I, this is subject to debate, see \citealt{siegel08}), traits that are associated primarily with dSphs, not GCs. Also, some of these faint satellites, such as Ursa Major II and Coma Berenices, are observed to have mean metallicites that are significantly higher than would be expected for objects of their luminosity \citep{simon07}. It is important to point out that such studies are hampered by low-number statistics, but the results certainly indicate the existence of an interesting population of low surface brightness satellites within the Local Group. Similarly, in M31 a large population of  unusual extended clusters (hereafter ECs$^1$) has been uncovered in the outskirts of the halo that cannot easily be classified as GCs or dSphs \citep{huxor05,huxor08}. Objects analogous to these ECs have also been discovered around other nearby galaxies. \citet{stonkute08} have identified one such cluster in M33 and \citet{hwang05} found 3 remote ECs in NGC 6822. The luminosities ($-4>M_V>-8$) and half light radii (10--35~pc) of these objects places them in an intriguing position in size-luminosity parameter space, populating what was previously observed as a gulf between dSphs and classical GCs (see fig.~\ref{lsize}).

\footnotetext[1]{We emphasize that while we refer to our objects as ``extended clusters'' (ECs),  no judgment on the nature of these systems is implied by this terminology.} 

Constraining the nature and properties of these low luminosity objects is important to our understanding of galaxy evolution and hierarchical structure formation. If they are the tidally-stripped cores of dSphs, such as has been suggested for Milky Way cluster, $\omega$ Cen, and the M31 cluster, G1 (\citet{bekki03}; \citet{bekki04}; \citet{ideta04}; \citet{tsuchiya04}), they are relics of previous accretion events within their host galaxy.
If, however, they reside in the extreme tail of log-normal size distribution of GCs
they are sufficiently unusual to warrant further investigation.
A serious problem in determining the properties, such as mass and metallicity, of these systems is correctly identifying which of the observed stars are cluster members. With spectroscopic information, one can separate stars that are kinematically associated
with the cluster from those that merely lie coincident with its Colour-Magnitude Diagram (CMD), and allow
us to derive dynamical constraints. In addition, the metallicity of individual stars can be measured from the spectra of the likely members via an inspection of the Near-Infrared CaII lines (CaT), providing a measure for \FeH\ that is independent
of the photometrically derived value. We can also use the central velocity dispersion
of the cluster to estimate the instantaneous mass of the system, assuming it is perfectly spherical
and in virial equilibrium (e.g. \citealt{illingworth76}, \citealt{richstone86}). One can then compare the M/L
ratio of ECs and other derived properties to those expected of GCs and dSphs in order to determine if they can be
reconciled entirely with the global properties of either, or if they are indeed members of another type of stellar 
association that bridges the gap between the two.

To this end, we have initiated a spectroscopic survey of the newly discovered extended clusters found in the halos of M31 and M33 using the DEIMOS spectrograph on Keck~II to derive radial velocities and metallicities of RGB stars. In this contribution, we discuss spectroscopic observations in EC4. \ec4\ was discovered within the CFHT-MegaCam survey (\citealt{huxor08}, Ibata et al.\ 2007) at a projected radius of 60~kpc (coordinates presented in Table 1). It has a half light radius of 33.7 pc and an absolute magnitude of M$_V$=-6.6 \citep{mackey06}, placing it in an usual position in terms of its size and luminosity as can be seen in Fig.~\ref{lsize}.

\section{Observations and analysis}

\begin{figure*}
\begin{center}
\includegraphics[angle=0,width=0.43\hsize]{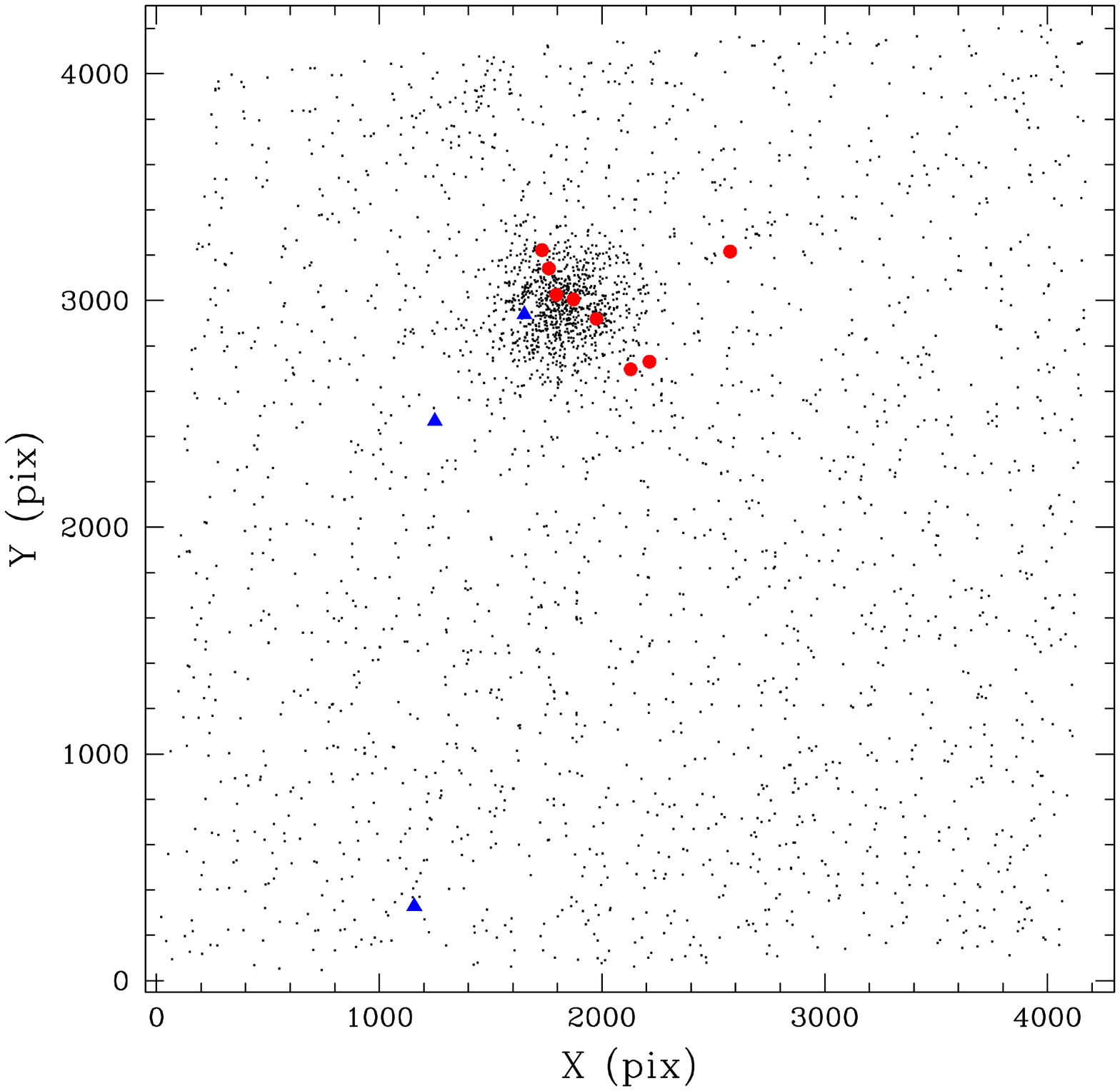}
\includegraphics[angle=0,width=0.43\hsize]{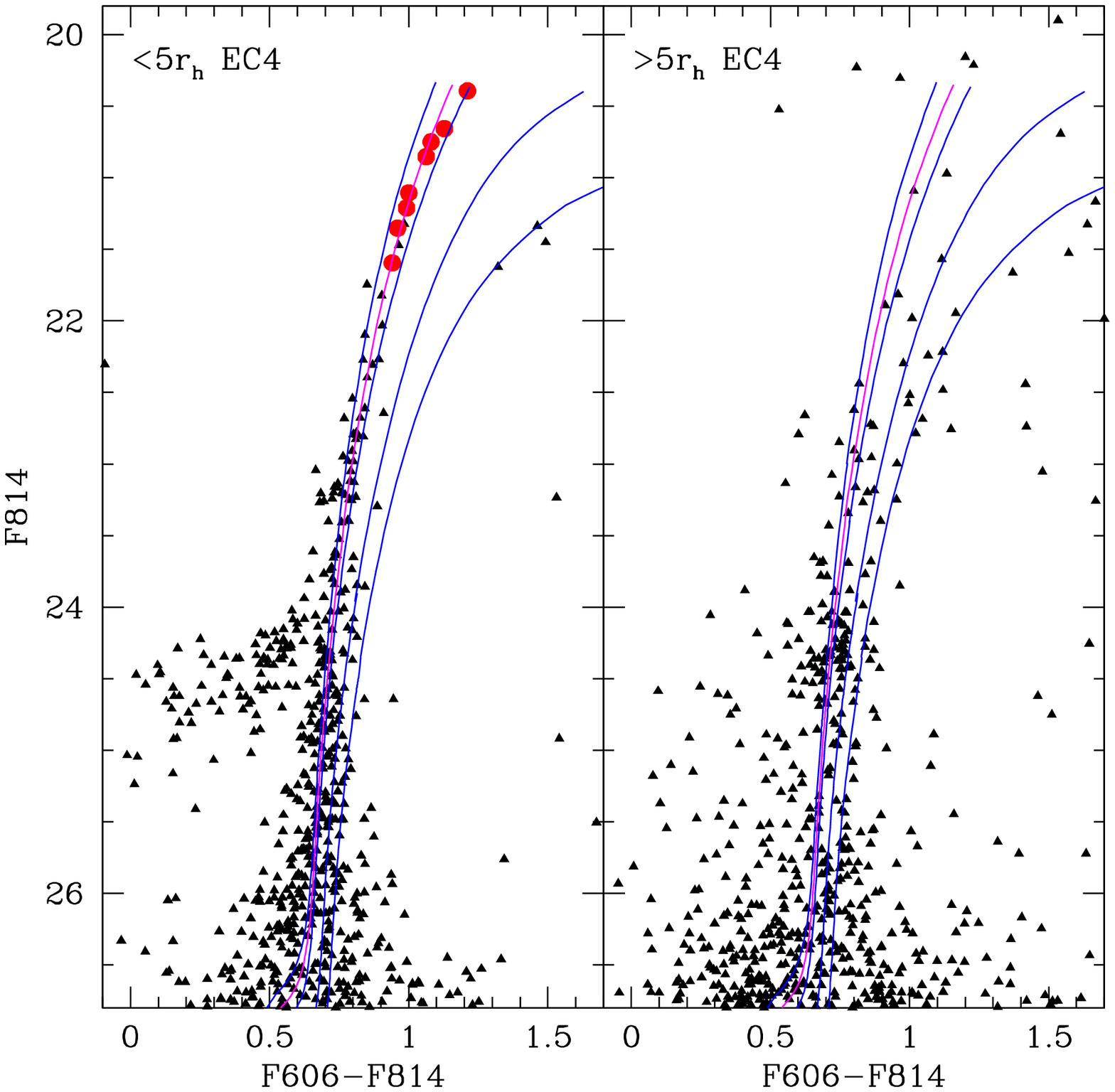}
\includegraphics[angle=0,width=0.43\hsize]{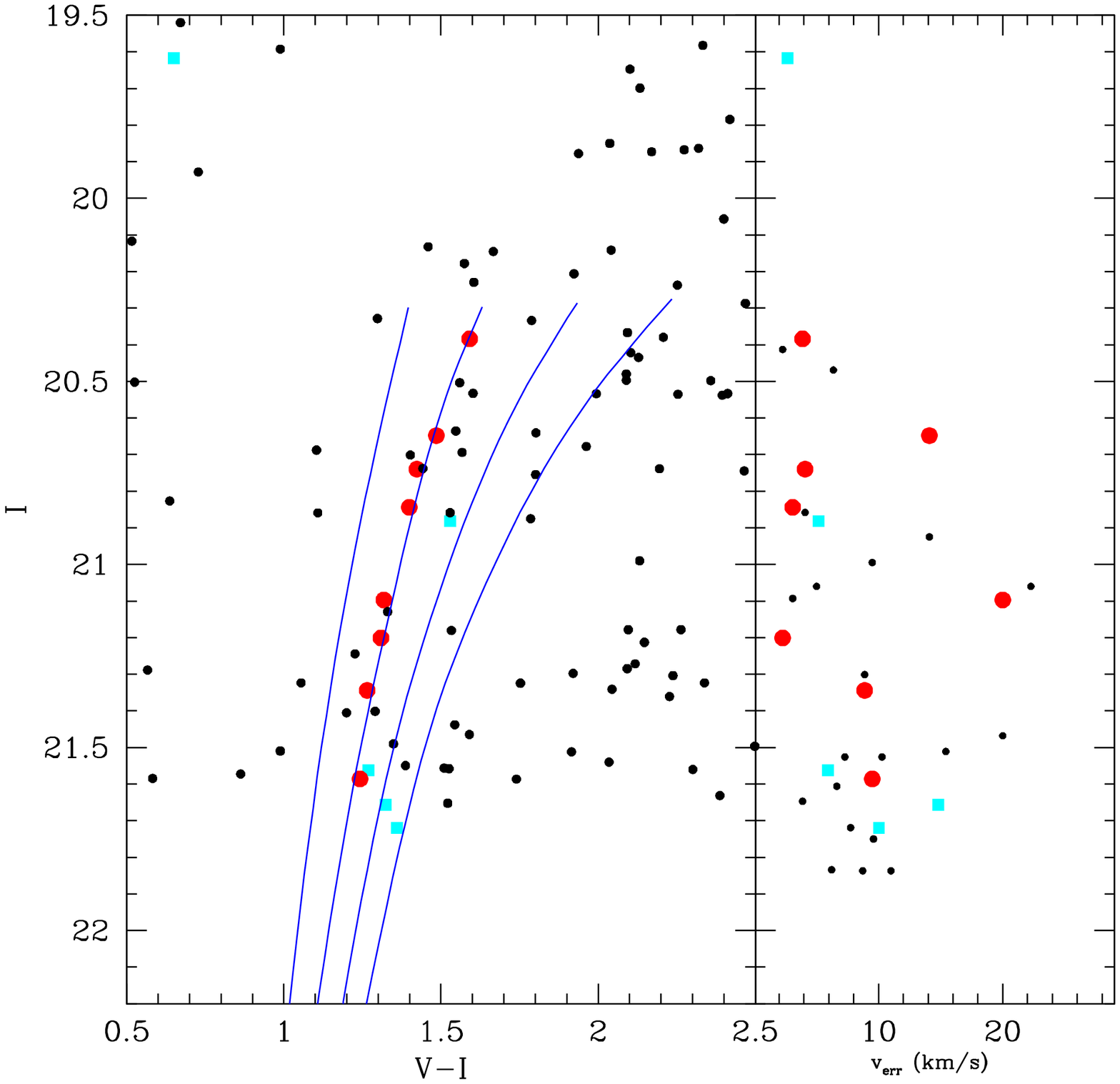}
\includegraphics[angle=0,width=0.43\hsize]{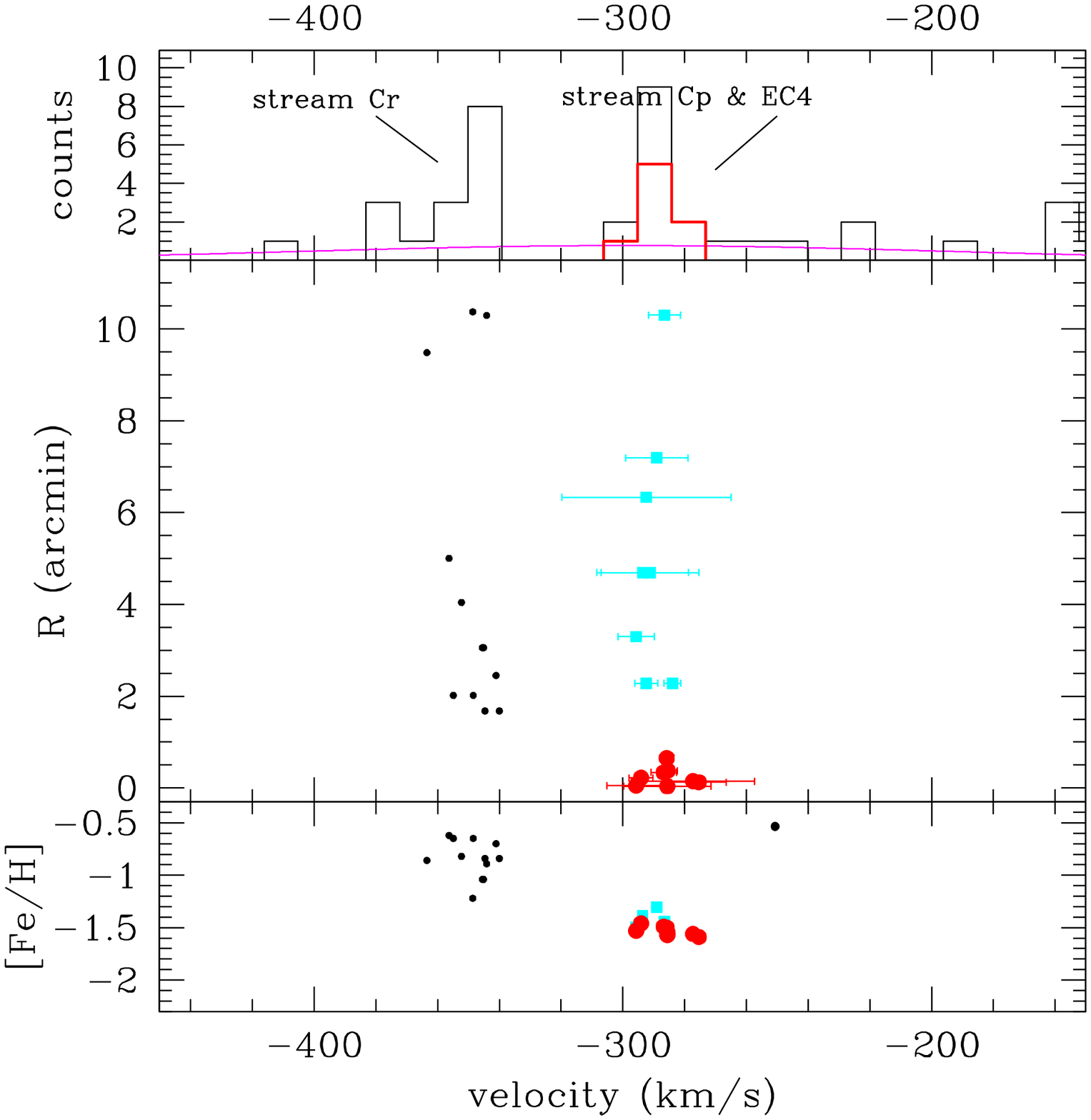}
\caption{
{\bf Top left panel: }Position plot of stars in the HST imagery of the \ec4\ region, 
highlighting the spectroscopic
candidates from our Keck survey. Circles indicate plausible EC4 candidates whilst the triangle represent likely members of \streamcp.  
{\bf Top right panel: }HST CMD of all objects within 0.8' (right) and outside of (left) $\sim$5$r_h$ of EC4. Our spectroscopic targets are highlighted with red circles. Isochrones overlaid are from the \citet{dart08} set, and correspond to (from left to right) -2.0, -1.8 -1.5, -1 and -0.75. These isochrones have been shifted to the
average distance modulus of EC4, 24.31 (Mackey et al.\ 2006). The isochrone at [Fe/H]=-1.8 provides the closest match for the likely EC4 stars in the left panel, and in the right panel while the more metal-rich \streamcr\ component can be clearly seen at around [Fe/H]=-1.0, we also see an RGB component persisting at [Fe/H]=-1.8, which we attribute to \streamcp.
{\bf Bottom left panel:}
 CFHT-MegaCam CMD and radial velocity uncertainties 
of the observed stars in the \streamcr/\streamcp\ fields.
Stars likely belonging to \ec4\ (red filled circles) and \streamcp\ (blue filled squares) are highlighted.
The isochrones are also from the \citet{dart08} set, and correspond to (from left to right) -2.0, -1.5, -1.25 and -1.0.   
{\bf Bottom right panel:}
The velocities of observed stars in the F25/F26 fields are shown in the top panel 
as a histogram, with EC4 member stars highlighted as a heavy histogram.
The stellar halo velocity dispersion ($\sigma_v$=125~kms$^{-1}$) 
from Chapman et al.\ (2006) is 
shown normalized to the expected 9 halo stars at this position from 
Ibata et al.\ (2007).
To differentiate EC4 stars from the field, we additionally plot
the velocities against their radius from the \ec4\ center in the centre panel,
referencing the symbols to the CMD plot.
In the lower panel, photometrically derived [Fe/H] from CFHT-MegaCam is shown as a function
of radial velocity,  again referenced in symbol type to the CMD plot.
}
\label{cmdvels}
\end{center}
\end{figure*}

The spectroscopic fields discussed in this paper are located at a projected distance of $\sim$60~kpc from the center of M31. The fields sit within \streamc presented in the Ibata et al.\ (2007)  M31 extended halo analysis,

Multi-object spectroscopic observations with the Keck-II telescope and 
DEIMOS (Davis et al.\ 2003) 
were obtained in photometric conditions with  $\sim0.8''$ seeing in Sept.\ 2004.
Target stars were chosen by colour/magnitude selection as described
in Ibata et al.\ (2005), first selecting likely RGB stars in M31 over all
metallicities, and filling the remaining mask slits with any other suitable stellar objects in the field.
Two spectroscopic masks (F25 and F26 from the table in
Chapman et al.\ 2006) targeted the
field of the extended cluster \ec4\ \citep{huxor08}.
A combined total of 212 independent stars in both masks
were observed in standard DEIMOS slit-mask mode (Davis et al.\ 2003) 
using the high resolution 1200 line/mm grating, and 1$''$ width slitlets.
Ten of these target stars were specifically selected from HST photometry of
\ec4\ to lie within the cluster. The top left panel of Figure~\ref{cmdvels} shows the map of HST stars derived from imaging in Mackey et al.\ (2006), and depicts the position of our spectroscopic candidates within the cluster.
Our instrumental setting 
covered the observed wavelength range from $\sim7000$--9800\AA.
Exposure time was 60 min, split into 20 min integrations. 
The DEIMOS-DEEP2 pipeline (Newman et al.\ 2004) 
designed to reduce data of this type accomplishes tasks of
debiassing, flat-fielding, extracting, wavelength-calibrating and
sky-subtracting the spectra.

The  radial velocities  of the  stars in all these fields were
then  measured with  respect  to spectra  of  standard stars  observed
during the observing runs. By fitting the peak of the cross-correlation
function, an estimate of the radial velocity accuracy was obtained for
each  radial velocity  measurement.  The accuracy  of  these data,
as estimated from the CaT cross-correlation,
varies with magnitude, having uncertainties  of
$<10\kms$ for most of the stars. The CMDs, velocity errors, velocity histograms, and metallicities for these fields are shown in Figure~\ref{cmdvels}.

We address Galactic contamination in our spectroscopically identified stars in a manner similar to Koch et al.\ (2008), using a combination of radial velocity and the equivalent width (EW) of  {Na\sc i}$_{\lambda8183,8195}$ which is
sensitive to surface gravity, and is accordingly very weak in M31
RGB star spectra, but can be strong in Galactic dwarfs (Schiavon et al.\ 1997). 
The Besan\c{c}on Galaxy model \citep{robin04} predicts that few contaminating foreground dwarfs  with v$_{r}<-160$~kms$^{-1}$ (as shown in \citealt{ibata05}) are present in a standard DEIMOS field (16.7' x 5'), and the chance of contamination within the small region of EC4 is almost nil. This is further supported by the fact that very few RGB candidates at these velocities show any
significant {Na\sc i}$_{\lambda8183,8195}$ absorption lines,
whereas stars with $-160$ to $0$~kms$^{-1}$
velocity show strong {Na\sc i} absorption on average, consistent with the findings of
Guhathakurta et al.\ (2006), Gilbert et al.\ (2006) and Koch et al.\ (2008).
For this study, we ignore all stars with v$_{r}>-160$~kms$^{-1}$, and we remove any stars from our sample which have a summed EW({Na\sc i}$_{\lambda8183,8195}$)$>0.8$   in the velocity range v$_{r}<-160$~kms$^{-1}$.

 In Fig.~\ref{cmdvels}, we present the CFHT MegaCam colours of all stars within our two DEIMOS fields, transformed into Landolt V and I using a two stage process detailed in \citet{ibata07} and \citet{mcconnachie04b}. Ten stars were selected to lie within \ec4\ by their colours and positions from the HST CMD, also shown in Fig.~\ref{cmdvels}. Of these ten, one was subsequently identified as a galaxy in the HST/ACS image, while another lies well off the cluster RGB, likely due to contamination in the ground-based CFHT-MegaCam imaging affecting the colour measurements. The remaining eight stars lie close to the cluster centre and fall along the top of the narrow RGB from the HST imagery. 

In analyzing the DEIMOS spectroscopy, one targeted star lay exactly on the 
\ec4\ CMD, but had a discrepant velocity from the others. Closer
examination of the spectrum revealed detections of the first and second CaT lines with a velocity of -277.3 kms$^{-1}$,
whereas the automated software pipeline derived a cross-correlation fit to larger skyline residuals.
We include this eighth star in the catalog as a viable member of \ec4, although we do not include it in subsequent analysis owing to its poorly defined velocity with respect to other member stars; we list the properties of all these stars in table~1.

\begin{figure*}
\begin{center}
\includegraphics[angle=0,width=0.43\hsize]{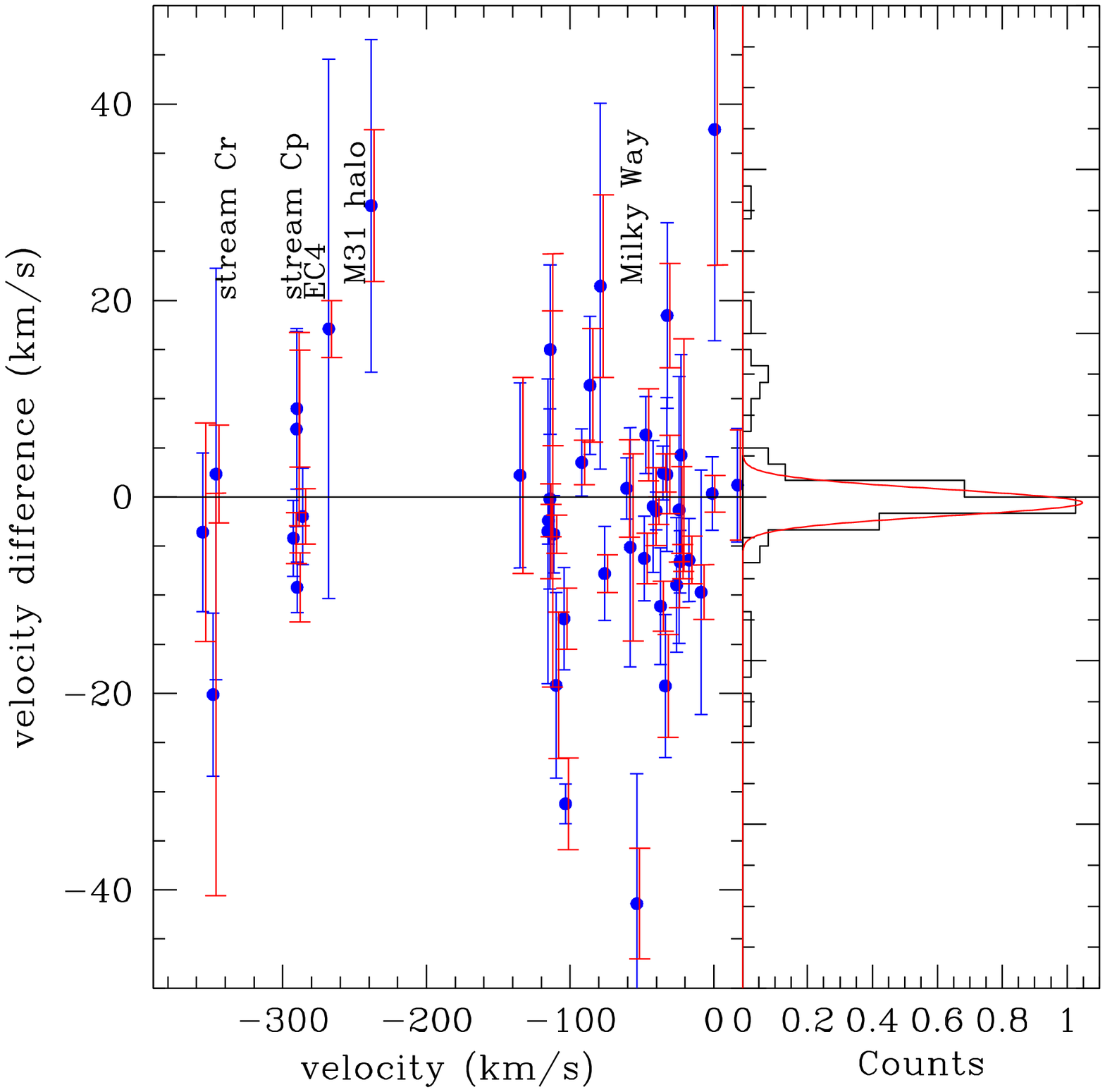}
\caption{ 
{\bf Left panel:} Velocity differences of stars lying in both fields F25 and F26.
Stars identified to velocity regions likely associated to 
\streamcr, \streamcp, \ec4, background M31 halo, and Milky Way foreground 
are identified.
Stars are shown at their velocities from mask F25, with offset error-bars
from mask F26. {\bf Right panel: }Histogram of velocity difference of stars lying in both fields F25 and F26. Normalizing this histogram gives a distribution that is reasonably fit by a Gaussian centred on -0.05 with $\sigma$=1.2.}
\label{veldiff}
\end{center}
\end{figure*}

\begin{figure*}
\begin{center}
\includegraphics[angle=270,width=0.655\hsize]{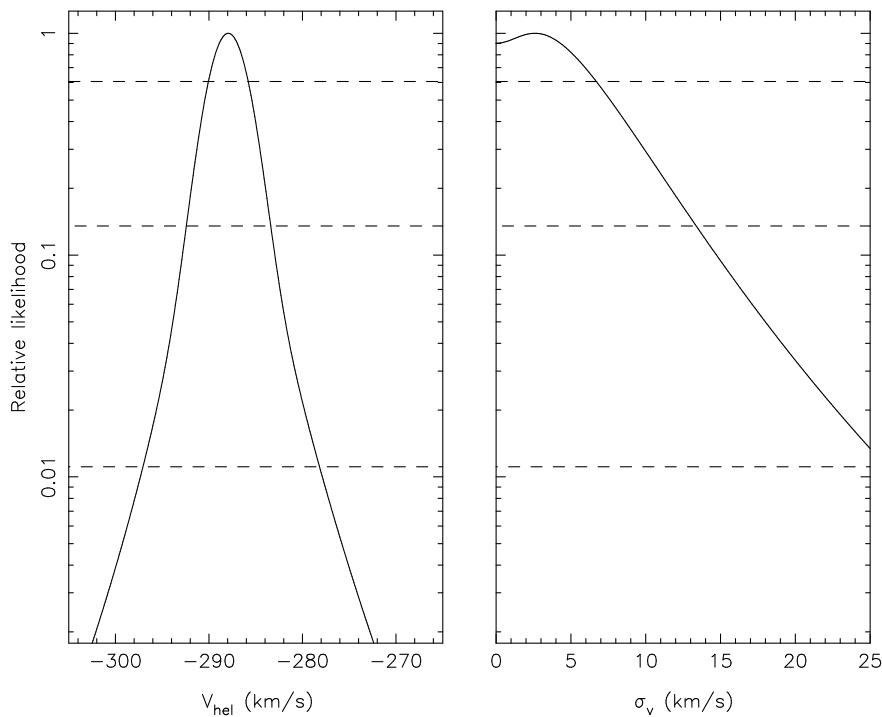}
\caption{
The kinematics of \ec4.
The relative likelihood distribution (taking into account the measurement errors) of v$_r$ and 
$\sigma_{v}$  for velocities of \ec4\ stars
when marginalizing with respect to the other parameter. The thin  
dashed lines correspond, from top to bottom, to the parameter range  
that contains 68.3\%, 95.4\% and 99.73\% of the probability distribution
(1, 2 and 3$\times$ $\sigma$ uncertainties),
showing that the velocity dispersion of EC4 is unresolved.
The peak values and $1\sigma$ range are $<v_r>=-287.9^{+1.9}_{-2.4}$ and $\sigma_{v}=2.7^{+4.2}_{-2.7}$.
}
\label{maxlike}
\end{center}
\end{figure*}

\subsection{Velocity accuracy}

Velocities for these EC4 stars were initially presented in \citet{chapman08}, but since publication, we have discovered a systematic effect that was introduced into some of our spectra from a crowding of slits within the EC4 region. This caused slits over some stars to be truncated resulting in a poor sky subtraction. The outcome was a broadening of the third Calcium Triplet line in a number of spectra, affecting the automated velocity cross correlations. We have since re-calibrated both our velocity and [Fe/H] calculations to include only the more reliable first and second triplet lines, giving us more robust kinematics for EC4 than presented previously.

While the cross-correlation errors for these CaT measurements suggest relatively small velocity errors, an independent check can be made on the 56 radial velocities of stars lying in spectroscopic masks of both fields F25 and F26. Velocity differences are shown in Fig.~\ref{veldiff}
 highlighting those corresponding to \streamc\ (3 stars), \streamd\ (2 stars), \ec4\ (3 stars), background halo (2 stars), and Milky Way foreground (46 stars).
Agreement between observing nights for these stars is generally 
found within the 1$\sigma$ errors of the radial velocity measurements, 
suggesting that no significant systematic errors are present from
instrumental setup night to night.
The dispersion in velocity differences is $\sim$6~kms$^{-1}$ for both the M31 sample and the Milky Way sample taken separately.
For these 56 stars, we have taken as the radial velocity the error-weighted average of the  measurement from the two nights. We also display in the right-hand panel of Fig.~\ref{veldiff} a histogram of the velocity differences for the overlap stars. Normalizing this histogram with respect to the individual errors of the stars within the field results in a distribution that is reasonably fit by a Gaussian centered on -0.05~kms with a dispersion of 1.2~kms, suggesting that instrumental errors night-to-night do not significantly affect derived values of velocity. We also investigated the correlation between velocity difference and slit position along the DEIMOS mask to see if there are any significant rotational effects present that could cause a misalignment of the slits from their target stars and introduce skewed velocity measurements across the mask. A slight trend is observed across the field, likely due to rotation of the mask,  but the effect is to impose a drift of $\sim0.2$\AA, corresponding to a velocity of 8.1 kms$^{-1}$ across the mask as a whole, or just 2 kms$^{-1}$ for the region of EC4, which is well within typical velocity errors.

\begin{figure*}
\begin{center}
\includegraphics[angle=0,width=0.8\hsize]{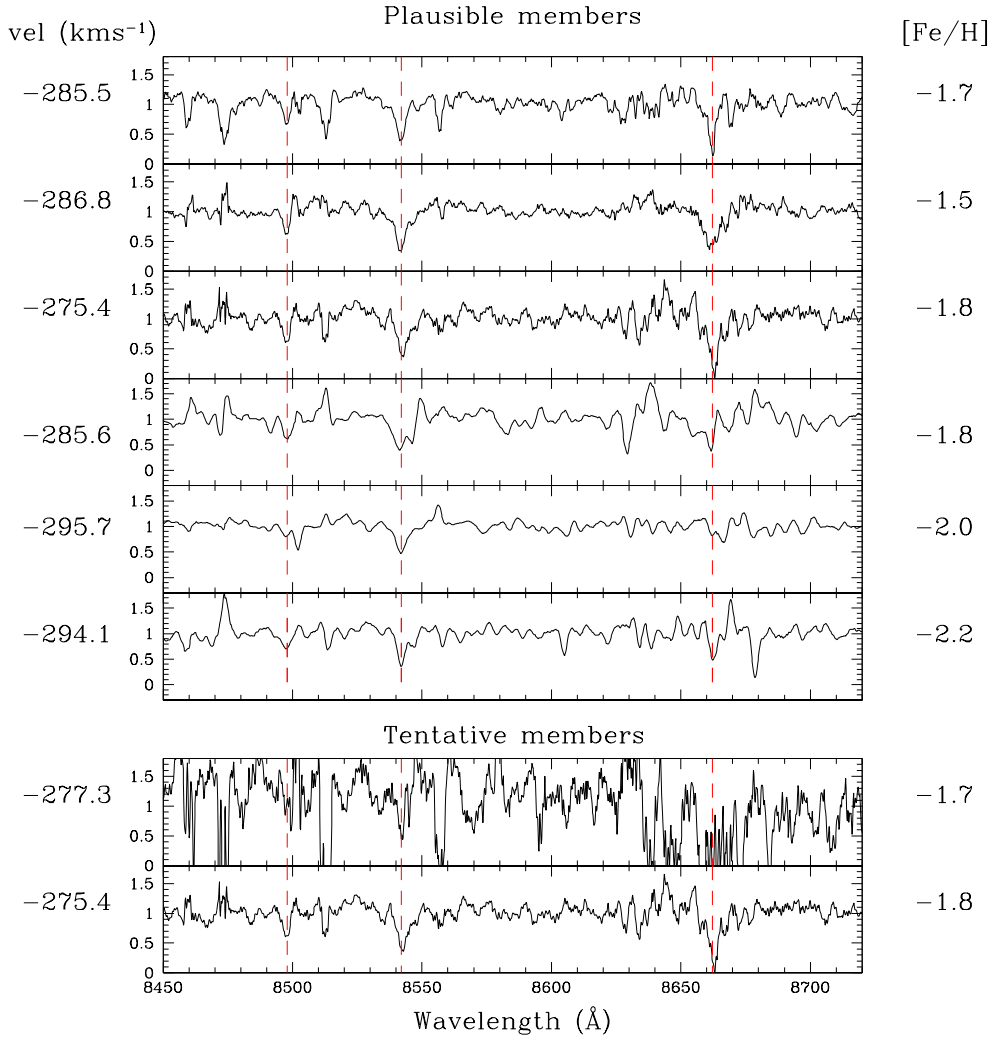}
\caption{
The individual spectra of the six plausible and the two tentative  EC4 members are displayed with their associated velocities and measurements of [Fe/H] (as measured from the first and second CaT lines). As can be seen, the third CaT line in the majority of the spectra is skewed as a result of poor sky subtraction.
}
\label{spectra}
\end{center}
\end{figure*}

\section{Results}

\subsection{Kinematics}

Figure~\ref{cmdvels} shows the
CFHT-MegaCam CMD  and radial velocity uncertainties
of the observed stars in the \ec4 fields. Stars likely belonging to \ec4\ are shown in red,
allowing metallicity comparison to the \citet{dart08} isochrones. Figure~\ref{cmdvels} displays the stars as a velocity histogram, highlighting the probable EC4 member stars as defined by their velocities. The kinematics for EC4 are strikingly similar to those of \streamcp as discussed in \citet{chapman08}, making it difficult to differentiate between the two objects on velocity alone. However, owing to the low density of \streamcp stars within the 83.5 arcmin$^2$ DEIMOS field (5-7 stars associated to \streamcp by \citet{chapman08}), it is unlikely that a contaminating stream star would be found within the $\sim0.8'$ tidal radius of the cluster and hence we can use the radial distance from the centre of the cluster as a discriminant between the two populations. This is also plotted in Fig.~\ref{cmdvels}.  Finally photometrically derived metallicities, [Fe/H] = $\log(Z/Z_\odot)$, 
are computed for the stars by interpolating between 10\,Gyr old 
Dartmouth isochrones \citep{dart08}. The fixed age and $\alpha$-abundance of +0.4 adopted for metallicity comparison to the fiducials 
will of course introduce a systematic uncertainty if significant variations are present. 
These photometric metallicities are plotted in Fig.~\ref{cmdvels} 
as a function of radial velocity.

Figure~\ref{cmdvels} shows that eight of the stars targeted in the cluster, 
\ec4, can be kinematically identified with a distribution of 
velocities
centered at v$_{r}\sim$-285kms$^{-1}$.  We exclude one of these from the following kinematic analysis as the noise in the spectrum is too high for us to ascertain a reliable velocity for the star. Of the remaining seven candidates, six lie within 2 core radii of the cluster centre and therefore are the most plausible \ec4 members. The seventh star lies further out at $\sim4$ core radii, a distance within which 99\% of the light of the cluster is contained, making its membership uncertain. For this reason, we omit this star in the following analysis also. The individual spectra for both plausible and tentative members are shown in Fig.~\ref{spectra}
 
The velocities for these six remaining stars are consistent with a v$_{r}$ = -287.9$^{+1.9}_{-2.4}$~kms$^{-1}$ and an instrumentally resolved $\sigma_{v,corr}=2.7^{+4.2}_{-2.7}$~kms$^{-1}$ after accounting for velocity measurement errors in the maximum-likelihood sense described previously
(see e.g., \citet{martin07} for details of the technique).
Figure~\ref{maxlike} shows the relative likelihood distribution of v$_r$ and $\sigma_{v}$  
when marginalizing with respect to the other parameter 
(the quoted uncertainties correspond to region containing  
the central 68.3\% of the distribution function). 
The parameter range 
that contains 68.3\%, 95.4\% and 99.73\% of the probability distribution
(the 1, 2 and 3$\times$ $\sigma$ uncertainties) are highlighted with dashed lines.
The plot shows that the velocity dispersion of EC4 is consistent with zero at the $1\sigma$ level.
 
Taking the measured value for the integrated light of EC4, $M_V =-6.6\pm0.1$ (Mackey et al.\ 2006), we can assess the dynamical mass of the cluster if we assume virial equilibrium.
The  mass to  light ratio  of a  simple spherically  symmetric stellar
system  of  central  surface  brightness $\Sigma_0$,  half  brightness
radius  $r_{h}$, and  central velocity  dispersion $\sigma_0$  can be
estimated as:
\begin{equation}
M/L =  \eta {{9}\over{2 \pi G}}  {{\sigma^2_0}\over{ \Sigma_0 r_{h}}}
\end{equation}
\citep{richstone86}, where $\eta$ is  a dimensionless parameter which has
a value close to unity for many structural models.

Using the values of r$_{h}$=33.7$\pm5$ pc and r$_{c}$=23.8$\pm4$ pc in \citet{mackey06}, one can calculate a tidal radius, r$_{t}$=165$\pm24$ pc. In order to get a value for central surface brightness, one can simply integrate the King profile in two dimensions, giving an area for the cluster of 3496~pc$^2$. Using the distance modulus from Mackey et al. of $(m-M)_{0}=24.31\pm0.14$ and assuming an absolute solar magnitude of M$_V$=4.83, we calculate the luminosity of the cluster to be L=3.73$^{+0.4}_{-0.3}$x 10$^4\lsun$. Hence we derive a value of $\mu_0=10.7\pm0.1 \lsun\/~pc^{-2}$ or 23.8 mag/arcsec$^{-2}$ which is fainter than typically observed in GCs \citep{noyola06} but considerably brighter than would be expected for a dSph of similar absolute magnitude \citep{mcconnachie06}.  

We assume the velocity dispersion we measure for EC4 is representative of $\sigma_0$ as all six stars lie within 2 core radii. This assumption would cause us, if anything, to underestimate $\sigma_{0}$ as in models of bound stellar systems, like those assumed in the core-fitting method, the  velocity  dispersion decreases with  distance, as our \ec4\ sample does marginally within the errors. If this is the case, the result would be that we would underestimate the M/L ratio of \ec4 also.

Assuming  that $\sigma_v = \sigma_0$, we find M/L$= 6.7^{+15}_{-6.7}\msun/\lsun$ (with errors denoting the 1$\sigma$ bound). Comparing this result with values of dSphs of similar luminosity, we find it unlikely that EC4 contains a sizable dark matter contribution. It is also unlikely that EC4 is the central remnant of a dSph that has lost the majority of its mass via tidal interaction with M31 or surrounding substructure, as the models of \citet{penarrubia08} show that severe mass loss from a dwarf spheroidal by means of tidal stripping actually {\bf increases} the value of M/L. This is because the tightly bound central dark matter ``cusp'' of the dwarf is more resilient to tidal disruption than the stellar component of the galaxy, causing the object to become more dark matter dominated. Another possibility is that EC4 is a stellar cluster undergoing tidal stripping, causing it to have an inflated mass-to-light ratio in comparison to more `typical' GCs (though we note that the M/L of EC4 is consistent with this population within its 1$\sigma$ errors). Significant tidal debris from the cluster is not evident in the HST photometry of \citet{mackey06}, though this could be due in part to the faintness of the cluster.

\begin{figure*}
\begin{center}
\includegraphics[angle=270,width=0.45\hsize]{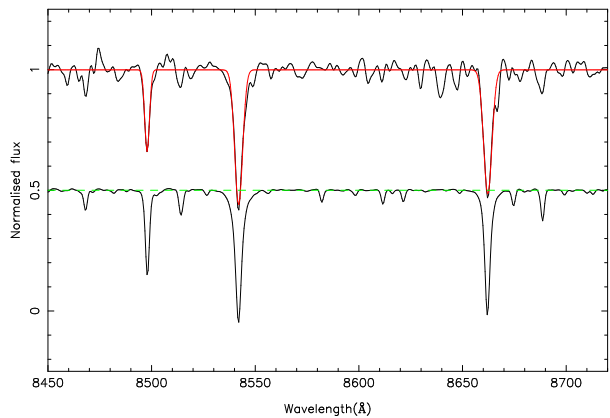}
\caption{
{\bf Top spectrum:} A straight average of reliable spectra plotted after normalizing by
continuum and smoothing using a Gaussian. Over-plotted in red are the 3 smoothed Gaussian fits to the CaT lines.
{\bf Lower spectrum:} For comparison, a continuum-normalized model atmosphere
spectrum from Munari et al. 2005 is shown, re-sampled to R=6000 with log(g)=1.0,
[Fe/H] = -1.5, T$_{eff}$ = 4500K and shifted down by 0.5 for ease of reference
}
\label{avspec}
\end{center}
\end{figure*}

\subsection{Measurements of \FeH}

 The metallicity of the cluster can be calculated using the spectroscopic data obtained for each member star from the equivalent widths of the CaT lines (as described in \citealt{ibata05}). The spectra for the six plausible members (plus the two tentative members) is shown in  Figure~\ref{spectra}, where velocities and individual metallicities for each candidate are quoted also. The spectroscopic metallicity of the six most probable member stars range from \FeH=-1.5 to \FeH=-2.2. It is important to note that the errors on the metallicities of each individual star are quite high (of order $\sim0.4$ dex) owing to the signal-to-noise of the data.

In Fig~\ref{avspec} we display the combined spectra for these stars after normalizing by
continuum and smoothed with a gaussian of instrumental resolution. Over-plotted in red are the 3 Gaussian fits to the CaT lines (also smoothed
to correspond to data). The lower plot is shown for comparison and represents a continuum-normalized model atmosphere
spectrum from Munari et al. 2005, re-sampled to R=6000 with log(g)=1.0,
[Fe/H] = -1.5, T$_{eff}$ = 4500K and shifted down by 0.5 for ease of reference. Using the combined spectra, an average metallicity of \FeH=-1.6$\pm0.15$ is determined for the cluster. The errors in this value correspond to the uncertainties in the measurements of the EWs of each CaT line, using the method described in \citet{battaglia08}. In a previous paper by \citet{mackey06}, a value of [Fe/H]= -1.84$\pm$0.20 is derived for \ec4 from HST photometry, which agrees well with our result within the quoted errors.

It is also interesting to note that 3 FeI lines (at $\sim$8467, $\sim$8515 and $\sim$8687\AA) can be seen in our summed spectra that appear to have similar EWs as their counterparts in the template spectrum, implying that the model is a good fit in surface gravity and temperature space to our data. The FeI lines could also allow us to estimate chemical abundance ratios with better spectra in the future (see \citealt{koch08,kirby08}). 

Photometric metallicities for the most likely members of EC4 were also determined, shown in the lower right panel of Fig.~\ref{cmdvels}, using isochrones from the Dartmouth stellar evoultion database \citep{dart08}. We selected isochrones with an age of 10 Gyr (as this was quoted as the lower limit for the age estimate for EC4 in \citealt{mackey06}), and a value of [$\alpha$/Fe]=+0.4 as GCs are typically observed with $\alpha$-abundances between +0.2 and +0.4. Then, using data from the MegaCam survey we interpolated between these isochrones to determine the photometric metallicity for each member star. Using this method, the mean metallicity for the cluster was determined to be \FeH=-1.5 with a dispersion of $\pm0.1$, which is more metal rich than the value of -1.84$\pm0.20$ found by \citet{mackey06}. However, in their paper, they used globular fiducials to obtain metallicities whereas we have used theoretical isochrones, which could expain this difference. Our choice of age (10 Gyr) is also younger than typical GCs, which could cause us to over-estimate the metallicity. This discrepancy is discussed in greater detail in the next section.

\subsection{Association with \streamcp}

From its projected position in the halo of M31, EC4 appears to sit within the tidal stream structure, \streamc, first reported in \citet{ibata07}. The work of \citet{chapman08} and Richardson et al. (2009) has since shown that this stream represents two superposed structures, a metal rich \streamcr\ and a metal poor \streamcp. It therefore seems natural to ask whether EC4 is related to either of these structures and, if so, what their relationship is?

From a comparison of their kinematics we find that, whilst \streamcr\ is the denser of these two streams in the region of EC4 ($\sim$3 times the density of \streamcp), it does not seem likely that it is related to the cluster by either its velocity (v$_r$=-349.5kms$^{-1}$ vs. -287.9kms$^{-1}$ for EC4) or its metallicity ([Fe/H]=-0.7 vs. -1.6 for EC4). However, from the 5 stars that were unambigously identified as members of \streamcp, Chapman et al. determine a systemic velocity for the structure of v$_r$= -285.0$^{+1.7}_{-1.8}$kms$^{-1}$, which is very similar to the velocity of -287.9kms$^{-1}$ that we derive here for EC4. They also found a spectroscopic metallicity for \streamcp\ of [Fe/H]=-1.3 which, taken at face value, is more metal-rich than the value of [Fe/H]=-1.6 that we find for EC4. However, on a closer inspection of the data used to calculate the metallicity for the stream, we note that the same broadening of the third CaT line that we see in the EC4 stars, is also observed in the \streamcp\ spectra, which will have resulted in an over-estimate of [Fe/H] for the stream. To correct this, we recalculate the metallicity for the stream using the first and second lines only and find [Fe/H]=-1.6$\pm0.2$, which is once again in remarkable agreement with our EC4 result.  

Photometrically, as stated in the previous section, we find a mean metallicity  for EC4 of [Fe/H]=-1.5$\pm0.1$. Using the same set of isochrones and applying them to the \streamcp\ stars we measure  a mean [Fe/H]=-1.4 $\pm0.1$, again in reasonable agreement. However given the errors from crowding and contamination within the CFHT data and the discrepancies between value of -1.5 we calculate here for EC4 and that of -1.84$\pm0.20$ calculated by \citet{mackey06}, it is important for us to assess whether the HST data shows a similar agreement between the two populations. As the spectroscopic \streamcp\ members were not observed in the HST pointing, we are unable to directly compare them to the EC4 members in the HST reference frame. Instead, we separate the EC4 stars within the HST field from the surrounding halo region by imposing a radial cut of 0.8', which is beyond the tidal radius of EC4, and compare the CMDs for both regions. This is shown in the top right panel of Fig~\ref{cmdvels}, where we display the HST CMDs with Dartmouth isochrones overlaid in the F606W and F814W filters, again with an age of 10 Gyr and [$\alpha$/Fe]=+0.4, at metallicities of [Fe/H]=-2.0, -1.8, -1.5, -1.0 and -0.75. As can be seen, the isochrone with [Fe/H]=-1.8 is the best match to the EC4 data (left panel). Turning our attention to the surrounding field (shown in the right hand panel), we can see the more metal rich \streamcr\ component quite clearly, but we also see a population persisting at [Fe/H]=-1.8. Given the distance of these stars from EC4, it is unlikely that they are bound members of the cluster, and hence we attribute this population to \streamcp, showing again a good agreement in the photometric properties of both objects.

The above findings strongly suggest that EC4 and \streamcp\ are somehow related to one another, and it is worth considering whether these five stream stars (and consequently, the surounding HST field stars) could be far-flung members of \ec4 itself, thereby simply representing rare or tidally disrupted members of an otherwise isolated cluster (without an associated tidal stream). These stars lie at large radii from \ec4 (2 to 10~arcmin, or $\sim0.5- 2$~kpc), chosen by chance in the spectroscopic masks, and are very unlikely to be members of \ec4\ itself. They would represent 15-74 core radii of \ec4 (30~pc) and therefore cannot belong to the cluster unless it is strongly disrupted, or is in fact only the cold core component of a more diffuse dwarf galaxy with an outer second component of stars. Owing to its low value of M/L (as discussed above), it is unlikely that this is the case. It is more probable that EC4 and the \streamcp\ stars both constitute tidally-stripped members of the as yet undiscovered stream progenitor.

\section{Discussion}

\subsection{ \ec4\ in the context of the dwarf spheroidal M/L to L relation}

\begin{figure*}
\begin{center}
\includegraphics[angle=0,width=0.395\hsize]{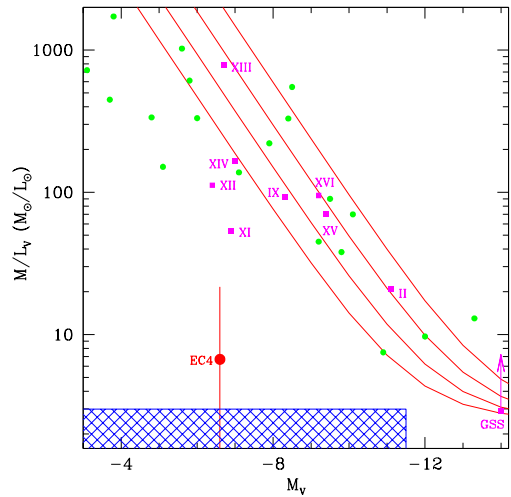}
\caption{
Comparison of the mass-to-light ratio (M/L) and luminosity of EC4 with known dwarf galaxies from
both the MW and M31 with central velocity dispersion estimates. The blue hatched region indicates the parameter space that GCs typically occupy.
All the estimates were made with the Richstone \& Tremaine (1986) equation presented as equation (1). \andxi, \andxii, and \andxiii\ (Collins et al. in prep.) are all shown as upper limits, since
their velocity dispersions are all unresolved by their measurements --
the 1$\sigma$ likelihood contour is used to set a tentative limit.
 The red lines are curves of constant dark matter halo mass (1, 2, 4, 8$\times$ 10$^7$ M$_\odot$ from bottom to top), assuming a stellar mass-to-light ratio of 2.5 M$_\odot$/L$_\odot$. It can be seen that whilst EC4 would need to be $\sim$100 times more massive than we have found here in order to consistent with dwarf galaxies of a similar luminosity, it is clearly distinct from the GC population also.
}
\label{mateo}
\end{center}
\end{figure*}

 Given that \ec4 encroaches on the parameter space between typical GCs and dSphs, it is interesting to compare \ec4 with other local group dSphs to see if its properties are consistent with this population. A comparison is made between EC4 and some of the known Milky Way and M31 satellites in Figure~\ref{mateo}, where the Mateo (1998) relation between M/L and the luminosity of dwarf galaxies in the Local Group has also been plotted, with the blue hatched region representing the parameter space typically occupied by GCs. 
Most of the newly discovered M31 dSphs are in agreement with the relation within reasonable errors. It is obvious that \ec4\ is not consistent with this relation, and would need to be of order $\sim$100 times more massive than we measure here for it to be comparable with other dwarfs of similar luminosity. This strongly implies that \ec4\ does not contain a significant quantity of dark matter, and is not a member of the dwarf spheroidal population. Instead we find that many of the attributes of EC4 are more consistent with those of a globular cluster without any sizable dark matter component. The half-light radius of 33~pc, though large for a cluster, is significantly smaller than what is typically observed in dSphs. There are three unusual, seemingly dark matter dominated objects where similar values of core radii to that of \ec4\ have been observed - Willman I, Segue I and Leo V \citep{willman06,belokurov06c,belokurov08} - though whether these associations represent dwarfs or clusters is also subject to debate (see e.g. \citealt{siegel08}). The objects are also vastly less luminous than \ec4 at M$_{V}$=-2.5, M$_{V}$=-3.0 and M$_{V}$=-4.3. and therefore they are not directly comparable. Another point to consider is that \ec4\ has a very narrow RGB, suggesting that it is composed of a single, probably old stellar population, without any evidence from the CMD of radially dependent differences between inner and outer parts \citep{mackey06}, which would suggest an extended star formation history. This is in contrast to dSphs where such trends amongst the population are typically observed, though there are a few exceptions to this (e.g. Ursa Major I, \citealt{okamoto08}). Finally, while it is conceivable that this object is associated with \streamcp, several other ECs with similar sizes and luminosities to EC4 are observed, both in M31, most notably ECs 1, 2 and 3 \citep{huxor05,mackey06,huxor08}, and M33 \citep{stonkute08}, that do not appear to be associated with stellar streams and are therefore not obvious candidates for dwarfs undergoing tidal disruption. On consideration of these results, we are led to conclude that \ec4 is most comparable to an extended globular cluster, possessing no measurable dark matter component.

\subsection{The origin of \ec4}

The location of \ec4\ in M31's halo, overlapping with the tangential streams `Cp' and `Cr', poses some interesting questions about the origins of the cluster. \ec4\ lies in a region where the metal-poor \streamcp\  has roughly 25\% the stellar density of the metal-rich \streamcr, 
although our spectroscopy reveals that \ec4\ is likely related to \streamcp\
with \streamcr\ overlapping only in projection. Could \streamcp\ actually be the debris from disrupted \ec4\ material?  The integrated luminosity of \streamcp\ is comparable to a faint M31 dwarf
galaxy like And\,XV or And\,XVI \citep{letarte09}, $M_V\sim-9.5$, which would suggest the baryonic matter mass loss of
\ec4\ ($M_V=-6.6$ Mackey et al.\ 2006) would be dramatically larger than its current {\it intact} mass.
No distortion of the \ec4\ isophotes is found in the HST imaging
of Mackey et al.\ (2006), although the faintness of \ec4\ means this
is unlikely to be a good test of ongoing mass loss or tidal distortion. Also, as noted in section 3, dSphs that have undergone severe tidal disruption are thought to increase their value of M/L and become more dark matter dominated. The low value of M/L=6.7$^{+15}_{-6.7}$ that we report here is therefore inconsistent with such an event. Using the results of \citet{penarrubia08}, we can estimate what the {\it initial} M/L of EC4 might have been if it had undergone heavy tidal stripping. They find that  for the most extreme cases of tidal disruption (mass loss $>99.7$\%), the velocity dispersions of dSphs decrease by a factor $\sim$4, the core radii are trimmed by a factor $\sim$3 and the surface density is decreased by factors up to $\sim$100. Applying these approximate corrections to our measured EC4 parameters gives an estimate for the initial M/L of less than 1, further weakening the case for EC4 being a member of the dSph population.

Regardless, it is likely that \streamcp\ and \ec4\ are at least related by their
kinematics and metallicities. If this is the case, then EC4 could represent an intact cluster of an as yet unidentified disrupting progenitor of \streamcp\ that has been stripped off its host and is now being carried along in its wake. Such a phenomenon has been observed in the MW, where several GCs have been identified within the stream of the currently disrupting Sagittarius dwarf that are thought to have been stripped from Sagittarius itself \citep{dinescu00,martinez02,cohen04,carraro07}. It is also possible that EC4 represents a cluster from M31's halo that is now being carried along within \streamcp.

There exists another possibility for the origins of \streamcp\ and \ec4. It has recently been suggested in \citet{fardal08} that by modeling the progenitor of the Giant Southern Stream (GSS) in M31 as a disk galaxy rather than a spherical, non-rotating galaxy as used in other models \citep{font06,fardal07}, many features of M31's halo, such as the Northeast and Western Shelves, can be reproduced. Preliminary results from these models show that with particular input configurations for the disk progenitor (specific combinations of disk mass, radius and orientation), the models can reproduce tangential arc structures, similar in metallicity and orientation about the minor axis to those observed by \citet{ibata07} and \citet{chapman08}. These arcs are the result of stripped debris from the rotating progenitor being left in different physical locations depending on the disks orientation with respect to M31. However these simulations do not reproduce the distinct values of [Fe/H] as seen in \streamcr\ and \streamcp. Another point to note is that initial results from modeling place these simulated stream-like structures much further from the centre of M31 than the observed streams, suggesting that their appearance in simulations may be coincidental. Further observational constraints and more in depth modeling is required to determine whether these arcs (and \ec4) could be related to the GSS. If this were the case, then it suggests that \ec4 could have been formed in a larger dwarf galaxy, and subsequently been stripped off as the progenitor was tidally disrupted by M31.

Of course the possibility exists that \ec4\ is not associated with the \streamc\ structure at all, that it merely overlaps the region in projection.
Without precise distance information, it is difficult to rule this out completely, though the striking similarities in the kinematics between the two objects makes such a conclusion seem unlikely.

\section{Conclusions}

In conclusion, we have conducted a Keck/DEIMOS survey of the extended cluster, EC4 in the halo of M31. We have measured the velocities and metallicities for 6 probable member stars and find values of v$_{r}$=-287.9$^{+1.9}_{-2.4}$~kms$^{-1}$, $\sigma_{v, corr}=2.7^{+4.2}_{-2.7}$~kms$^{-1}$, and [Fe/H]=-1.6$\pm0.15$ for the cluster.  The velocity dispersion of \ec4\ suggests that no sizable contribution of dark matter is present within \ec4\ as it has a M/L=6.7$^{+15}_{-6.7}$, although the results clearly carry large uncertainties based on only 6 stars. We find that EC4 is not comparable with typical M31 and MW dwarf galaxies, as it is not consistent with the well established M/L to L relation for dwarfs. It is more comparable with a GC whose large r$_{h}$ and M/L place it in the extreme tail of the log-normal size distribution of globulars, without a sizable dark matter component. 

EC4 lies in the region of \streamc, and its kinematics and metallicity are very similar to those of the more metal-poor component, \streamcp, suggesting that the two structures are related to one another. Owing to its current intact mass and M/L ratio, EC4 is unlikely to be the disrupting progenitor of the stream, but its strong kinematic resemblance suggests it is either a cluster that has been stripped from the progenitor, or a stellar system that is being carried along by the disrupted stream. 

``Extended Clusters'' possess some unusual and interesting properties. Whether they are  obscure globular clusters, tidally stripped dwarfs or a ``missing link'' between the two, they represent a relatively unstudied variant of stellar association. The work we have presented here indicates a need for further spectroscopic surveys of similar objects within M31 and other local group galaxies to be carried out, to see if results such as those found for EC4 are present throughout the EC population. It will be interesting to see if other ECs have associated streamy substructure, and to discover what role ECs play in the evolution of galaxies such as our own MW.

\section{ACKNOWLEDGMENTS}
MLMC acknowledges the award of an STFC studentship. SCC acknowledges NSERC for support.

\newcommand{\mnras}{MNRAS}
\newcommand{\pasa}{PASA}
\newcommand{\nat}{Nature}
\newcommand{\araa}{ARAA}
\newcommand{\aj}{AJ}
\newcommand{\apj}{ApJ}
\newcommand{\apjl}{ApJ}
\newcommand{\apjs}{ApJSupp}
\newcommand{\aap}{A\&A}
\newcommand{\aaps}{A\&ASupp}
\newcommand{\pasp}{PASP}

\bibliography{mnemonic,michelle}{}

\begin{thebibliography}{}

\bibitem[\protect\citeauthoryear{{Battaglia}, {Irwin}, {Tolstoy}, {Hill},
  {Helmi}, {Letarte} \& {Jablonka}}{{Battaglia} et~al.}{2008}]{battaglia08}
{Battaglia} G.,  {Irwin} M.,  {Tolstoy} E.,  {Hill} V.,  {Helmi} A.,  {Letarte}
  B.,    {Jablonka} P.,  2008, \mnras, 383, 183

\bibitem[\protect\citeauthoryear{{Bekki} \& {Chiba}}{{Bekki} \&
  {Chiba}}{2004}]{bekki04}
{Bekki} K.,  {Chiba} M.,  2004, \aap, 417, 437

\bibitem[\protect\citeauthoryear{{Bekki} \& {Freeman}}{{Bekki} \&
  {Freeman}}{2003}]{bekki03}
{Bekki} K.,  {Freeman} K.~C.,  2003, \mnras, 346, L11

\bibitem[\protect\citeauthoryear{{Belokurov} et~al.,}{{Belokurov}
  et~al.}{2007}]{belokurov07}
{Belokurov} V.,  et~al., 2007, \apj, 654, 897

\bibitem[\protect\citeauthoryear{{Belokurov}, {Evans}, {Irwin}, {Hewett} \&
  {Wilkinson}}{{Belokurov} et~al.}{2006}]{belokurov06c}
{Belokurov} V.,  {Evans} N.~W.,  {Irwin} M.~J.,  {Hewett} P.~C.,    {Wilkinson}
  M.~I.,  2006, \apjl, 637, L29

\bibitem[\protect\citeauthoryear{{Belokurov}, {Walker}, {Evans}, {Faria},
  {Gilmore}, {Irwin}, {Koposov}, {Mateo}, {Olszewski} \& {Zucker}}{{Belokurov}
  et~al.}{2008}]{belokurov08}
{Belokurov} V.,  {Walker} M.~G.,  {Evans} N.~W.,  {Faria} D.~C.,  {Gilmore} G.,
   {Irwin} M.~J.,  {Koposov} S.,  {Mateo} M.,  {Olszewski} E.,    {Zucker}
  D.~B.,  2008, \apjl, 686, L83

\bibitem[\protect\citeauthoryear{{Carraro}, {Zinn} \& {Moni Bidin}}{{Carraro}
  et~al.}{2007}]{carraro07}
{Carraro} G.,  {Zinn} R.,    {Moni Bidin} C.,  2007, \aap, 466, 181

\bibitem[\protect\citeauthoryear{{Chapman}, {Ibata}, {Irwin}, {Koch},
  {Letarte}, {Martin}, {Collins}, {Lewis}, {McConnachie}, {Pe{\~n}arrubia},
  {Rich}, {Trethewey}, {Ferguson}, {Huxor} \& {Tanvir}}{{Chapman}
  et~al.}{2008}]{chapman08}
{Chapman} S.~C.,  {Ibata} R.,  {Irwin} M.,  {Koch} A.,  {Letarte} B.,  {Martin}
  N.,  {Collins} M.,  {Lewis} G.~F.,  {McConnachie} A.,  {Pe{\~n}arrubia} J.,
  {Rich} R.~M.,  {Trethewey} D.,  {Ferguson} A.,  {Huxor} A.,    {Tanvir} N.,
  2008, \mnras, 390, 1437

\bibitem[\protect\citeauthoryear{{Chapman}, {Ibata}, {Lewis}, {Ferguson},
  {Irwin}, {McConnachie} \& {Tanvir}}{{Chapman} et~al.}{2005}]{chapman05}
{Chapman} S.~C.,  {Ibata} R.,  {Lewis} G.~F.,  {Ferguson} A.~M.~N.,  {Irwin}
  M.,  {McConnachie} A.,    {Tanvir} N.,  2005, \apjl, 632, L87

\bibitem[\protect\citeauthoryear{{Cohen}}{{Cohen}}{2004}]{cohen04}
{Cohen} J.~G.,  2004, \aj, 127, 1545

\bibitem[\protect\citeauthoryear{{Dinescu}, {Majewski}, {Girard} \&
  {Cudworth}}{{Dinescu} et~al.}{2000}]{dinescu00}
{Dinescu} D.~I.,  {Majewski} S.~R.,  {Girard} T.~M.,    {Cudworth} K.~M.,
  2000, \aj, 120, 1892

\bibitem[\protect\citeauthoryear{{Dotter}, {Chaboyer}, {Jevremovic}, {Kostov},
  {Baron} \& {Ferguson}}{{Dotter} et~al.}{2008}]{dart08}
{Dotter} A.,  {Chaboyer} B.,  {Jevremovic} D.,  {Kostov} V.,  {Baron} E.,
  {Ferguson} J.~W.,  2008, ArXiv e-prints, 0804.4473

\bibitem[\protect\citeauthoryear{{Fardal}, {Babul}, {Guhathakurta}, {Gilbert}
  \& {Dodge}}{{Fardal} et~al.}{2008}]{fardal08}
{Fardal} M.~A.,  {Babul} A.,  {Guhathakurta} P.,  {Gilbert} K.~M.,    {Dodge}
  C.,  2008, ArXiv e-prints, 0803.3476

\bibitem[\protect\citeauthoryear{{Fardal}, {Guhathakurta}, {Babul} \&
  {McConnachie}}{{Fardal} et~al.}{2007}]{fardal07}
{Fardal} M.~A.,  {Guhathakurta} P.,  {Babul} A.,    {McConnachie} A.~W.,  2007,
  \mnras, 380, 15

\bibitem[\protect\citeauthoryear{{Ferguson}, {Irwin}, {Ibata}, {Lewis} \&
  {Tanvir}}{{Ferguson} et~al.}{2002}]{ferguson02}
{Ferguson} A.~M.~N.,  {Irwin} M.~J.,  {Ibata} R.~A.,  {Lewis} G.~F.,
  {Tanvir} N.~R.,  2002, \aj, 124, 1452

\bibitem[\protect\citeauthoryear{{Font}, {Johnston}, {Guhathakurta}, {Majewski}
  \& {Rich}}{{Font} et~al.}{2006}]{font06}
{Font} A.~S.,  {Johnston} K.~V.,  {Guhathakurta} P.,  {Majewski} S.~R.,
  {Rich} R.~M.,  2006, \aj, 131, 1436

\bibitem[\protect\citeauthoryear{{Huxor}, {Tanvir}, {Ferguson}, {Irwin},
  {Ibata}, {Bridges} \& {Lewis}}{{Huxor} et~al.}{2008}]{huxor08}
{Huxor} A.~P.,  {Tanvir} N.~R.,  {Ferguson} A.~M.~N.,  {Irwin} M.~J.,  {Ibata}
  R.,  {Bridges} T.,    {Lewis} G.~F.,  2008, \mnras, 385, 1989

\bibitem[\protect\citeauthoryear{{Huxor}, {Tanvir}, {Irwin}, {Ibata},
  {Collett}, {Ferguson}, {Bridges} \& {Lewis}}{{Huxor} et~al.}{2005}]{huxor05}
{Huxor} A.~P.,  {Tanvir} N.~R.,  {Irwin} M.~J.,  {Ibata} R.,  {Collett} J.~L.,
  {Ferguson} A.~M.~N.,  {Bridges} T.,    {Lewis} G.~F.,  2005, \mnras, 360,
  1007

\bibitem[\protect\citeauthoryear{{Hwang}, {Lee}, {Lee}, {Park}, {Park}, {Park},
  {Sohn}, {Lee}, {Lee}, {Chun}, {Lee}, {Sohn}, {Yuk}, {Kim}, {Kim} \&
  {Han}}{{Hwang} et~al.}{2005}]{hwang05}
{Hwang} N.,  {Lee} M.~G.,  {Lee} J.~C.,  {Park} W.-K.,  {Park} H.~S.,  {Park}
  J.-H.,  {Sohn} S.~T.,  {Lee} S.-G.,  {Lee} H.~M.,  {Chun} M.-S.,  {Lee}
  Y.-W.,  {Sohn} Y.-J.,  {Yuk} I.-S.,  {Kim} S.~C.,  {Kim} H.-I.,    {Han} W.,
  2005, in {Jerjen} H.,  {Binggeli} B.,  eds, IAU Colloq. 198: Near-fields
  cosmology with dwarf elliptical galaxies {Discovery of remote star clusters
  in the halo of the Irregular galaxy NGC 6822}.
pp 257--258

\bibitem[\protect\citeauthoryear{{Ibata}, {Chapman}, {Ferguson}, {Lewis},
  {Irwin} \& {Tanvir}}{{Ibata} et~al.}{2005}]{ibata05}
{Ibata} R.,  {Chapman} S.,  {Ferguson} A.~M.~N.,  {Lewis} G.,  {Irwin} M.,
  {Tanvir} N.,  2005, \apj, 634, 287

\bibitem[\protect\citeauthoryear{{Ibata}, {Irwin}, {Lewis}, {Ferguson} \&
  {Tanvir}}{{Ibata} et~al.}{2001}]{ibata01b}
{Ibata} R.,  {Irwin} M.,  {Lewis} G.,  {Ferguson} A.~M.~N.,    {Tanvir} N.,
  2001, \nat, 412, 49

\bibitem[\protect\citeauthoryear{{Ibata}, {Martin}, {Irwin}, {Chapman},
  {Ferguson}, {Lewis} \& {McConnachie}}{{Ibata} et~al.}{2007}]{ibata07}
{Ibata} R.,  {Martin} N.~F.,  {Irwin} M.,  {Chapman} S.,  {Ferguson} A.~M.~N.,
  {Lewis} G.~F.,    {McConnachie} A.~W.,  2007, \apj, 671, 1591

\bibitem[\protect\citeauthoryear{{Ideta} \& {Makino}}{{Ideta} \&
  {Makino}}{2004}]{ideta04}
{Ideta} M.,  {Makino} J.,  2004, \apjl, 616, L107

\bibitem[\protect\citeauthoryear{{Illingworth}}{{Illingworth}}{1976}]{illingwo%
rth76}
{Illingworth} G.,  1976, \apj, 204, 73

\bibitem[\protect\citeauthoryear{{Irwin} \& {Hatzidimitriou}}{{Irwin} \&
  {Hatzidimitriou}}{1995}]{irwin95}
{Irwin} M.,  {Hatzidimitriou} D.,  1995, \mnras, 277, 1354

\bibitem[\protect\citeauthoryear{{Irwin} et~al.,}{{Irwin}
  et~al.}{2007}]{irwin07}
{Irwin} M.~J.,  et~al., 2007, \apjl, 656, L13

\bibitem[\protect\citeauthoryear{{Irwin}, {Ferguson}, {Huxor}, {Tanvir},
  {Ibata} \& {Lewis}}{{Irwin} et~al.}{2008}]{irwin08}
{Irwin} M.~J.,  {Ferguson} A.~M.~N.,  {Huxor} A.~P.,  {Tanvir} N.~R.,  {Ibata}
  R.~A.,    {Lewis} G.~F.,  2008, \apjl, 676, L17

\bibitem[\protect\citeauthoryear{{Kirby}, {Guhathakurta} \& {Sneden}}{{Kirby}
  et~al.}{2008}]{kirby08}
{Kirby} E.~N.,  {Guhathakurta} P.,    {Sneden} C.,  2008, \apj, 682, 1217

\bibitem[\protect\citeauthoryear{{Koch}, {Rich}, {Reitzel}, {Martin}, {Ibata},
  {Chapman}, {Majewski}, {Mori}, {Loh}, {Ostheimer} \& {Tanaka}}{{Koch}
  et~al.}{2008}]{koch08}
{Koch} A.,  {Rich} R.~M.,  {Reitzel} D.~B.,  {Martin} N.~F.,  {Ibata} R.~A.,
  {Chapman} S.~C.,  {Majewski} S.~R.,  {Mori} M.,  {Loh} Y.-S.,  {Ostheimer}
  J.~C.,    {Tanaka} M.,  2008, \apj, 689, 958

\bibitem[\protect\citeauthoryear{{Letarte}, {Chapman}, {Collins}, {Ibata},
  {Irwin}, {Ferguson}, {Lewis}, {Martin}, {McConnachie} \& {Tanvir}}{{Letarte}
  et~al.}{2009}]{letarte09}
{Letarte} B.,  {Chapman} S.~C.,  {Collins} M.,  {Ibata} R.~A.,  {Irwin} M.~J.,
  {Ferguson} A.~M.~N.,  {Lewis} G.~F.,  {Martin} N.,  {McConnachie} A.,
  {Tanvir} N.,  2009, ArXiv e-prints

\bibitem[\protect\citeauthoryear{{Mackey}, {Huxor}, {Ferguson}, {Tanvir},
  {Irwin}, {Ibata}, {Bridges}, {Johnson} \& {Lewis}}{{Mackey}
  et~al.}{2006}]{mackey06}
{Mackey} A.~D.,  {Huxor} A.,  {Ferguson} A.~M.~N.,  {Tanvir} N.~R.,  {Irwin}
  M.,  {Ibata} R.,  {Bridges} T.,  {Johnson} R.~A.,    {Lewis} G.,  2006,
  \apjl, 653, L105

\bibitem[\protect\citeauthoryear{{Mackey}, {Huxor}, {Ferguson}, {Tanvir},
  {Irwin}, {Ibata}, {Bridges}, {Johnson} \& {Lewis}}{{Mackey}
  et~al.}{2007}]{mackey07}
{Mackey} A.~D.,  {Huxor} A.,  {Ferguson} A.~M.~N.,  {Tanvir} N.~R.,  {Irwin}
  M.,  {Ibata} R.,  {Bridges} T.,  {Johnson} R.~A.,    {Lewis} G.,  2007,
  \apjl, 655, L85

\bibitem[\protect\citeauthoryear{{Martin}, {de Jong} \& {Rix}}{{Martin}
  et~al.}{2008}]{martin08}
{Martin} N.~F.,  {de Jong} J.~T.~A.,    {Rix} H.-W.,  2008, \apj, 684, 1075

\bibitem[\protect\citeauthoryear{{Martin}, {Ibata}, {Chapman}, {Irwin} \&
  {Lewis}}{{Martin} et~al.}{2007}]{martin07}
{Martin} N.~F.,  {Ibata} R.~A.,  {Chapman} S.~C.,  {Irwin} M.,    {Lewis}
  G.~F.,  2007, \mnras, 380, 281

\bibitem[\protect\citeauthoryear{{Martin}, {Ibata}, {Irwin}, {Chapman},
  {Lewis}, {Ferguson}, {Tanvir} \& {McConnachie}}{{Martin}
  et~al.}{2006}]{martin06}
{Martin} N.~F.,  {Ibata} R.~A.,  {Irwin} M.~J.,  {Chapman} S.,  {Lewis} G.~F.,
  {Ferguson} A.~M.~N.,  {Tanvir} N.,    {McConnachie} A.~W.,  2006, \mnras,
  371, 1983

\bibitem[\protect\citeauthoryear{{Mart{\'{\i}}nez-Delgado}, {Zinn}, {Carrera}
  \& {Gallart}}{{Mart{\'{\i}}nez-Delgado} et~al.}{2002}]{martinez02}
{Mart{\'{\i}}nez-Delgado} D.,  {Zinn} R.,  {Carrera} R.,    {Gallart} C.,
  2002, \apjl, 573, L19

\bibitem[\protect\citeauthoryear{{McConnachie}, {Huxor}, {Martin}, {Irwin},
  {Chapman}, {Fahlman}, {Ferguson}, {Ibata}, {Lewis}, {Richer} \&
  {Tanvir}}{{McConnachie} et~al.}{2008}]{mcconnachie08}
{McConnachie} A.,  {Huxor} A.,  {Martin} N.,  {Irwin} M.,  {Chapman} S.,
  {Fahlman} G.,  {Ferguson} A.,  {Ibata} R.,  {Lewis} G.,  {Richer} H.,
  {Tanvir} N.,  2008, ArXiv e-prints, 0806.3988

\bibitem[\protect\citeauthoryear{{McConnachie} \& {Irwin}}{{McConnachie} \&
  {Irwin}}{2006}]{mcconnachie06}
{McConnachie} A.~W.,  {Irwin} M.~J.,  2006, \mnras, 365, 1263

\bibitem[\protect\citeauthoryear{{McConnachie}, {Irwin}, {Ferguson}, {Ibata},
  {Lewis} \& {Tanvir}}{{McConnachie} et~al.}{2004}]{mcconnachie04b}
{McConnachie} A.~W.,  {Irwin} M.~J.,  {Ferguson} A.~M.~N.,  {Ibata} R.~A.,
  {Lewis} G.~F.,    {Tanvir} N.,  2004, \mnras, 350, 243

\bibitem[\protect\citeauthoryear{{McConnachie}, {Irwin}, {Ferguson}, {Ibata},
  {Lewis} \& {Tanvir}}{{McConnachie} et~al.}{2005}]{mcconnachie05a}
{McConnachie} A.~W.,  {Irwin} M.~J.,  {Ferguson} A.~M.~N.,  {Ibata} R.~A.,
  {Lewis} G.~F.,    {Tanvir} N.,  2005, \mnras, 356, 979

\bibitem[\protect\citeauthoryear{{Noyola} \& {Gebhardt}}{{Noyola} \&
  {Gebhardt}}{2006}]{noyola06}
{Noyola} E.,  {Gebhardt} K.,  2006, \aj, 132, 447

\bibitem[\protect\citeauthoryear{{Okamoto}, {Arimoto}, {Yamada} \&
  {Onodera}}{{Okamoto} et~al.}{2008}]{okamoto08}
{Okamoto} S.,  {Arimoto} N.,  {Yamada} Y.,    {Onodera} M.,  2008, ArXiv
  e-prints, 0804.2976

\bibitem[\protect\citeauthoryear{{Pe{\~n}arrubia}, {McConnachie} \&
  {Navarro}}{{Pe{\~n}arrubia} et~al.}{2008}]{penarrubia08}
{Pe{\~n}arrubia} J.,  {McConnachie} A.~W.,    {Navarro} J.~F.,  2008, \apj,
  672, 904

\bibitem[\protect\citeauthoryear{{Richstone} \& {Tremaine}}{{Richstone} \&
  {Tremaine}}{1986}]{richstone86}
{Richstone} D.~O.,  {Tremaine} S.,  1986, \aj, 92, 72

\bibitem[\protect\citeauthoryear{{Robin}, {Reyl{\'e}}, {Derri{\`e}re} \&
  {Picaud}}{{Robin} et~al.}{2004}]{robin04}
{Robin} A.~C.,  {Reyl{\'e}} C.,  {Derri{\`e}re} S.,    {Picaud} S.,  2004,
  \aap, 416, 157

\bibitem[\protect\citeauthoryear{{Siegel}, {Shetrone} \& {Irwin}}{{Siegel}
  et~al.}{2008}]{siegel08}
{Siegel} M.~H.,  {Shetrone} M.~D.,    {Irwin} M.,  2008, \aj, 135, 2084

\bibitem[\protect\citeauthoryear{{Simon} \& {Geha}}{{Simon} \&
  {Geha}}{2007}]{simon07}
{Simon} J.~D.,  {Geha} M.,  2007, \apj, 670, 313

\bibitem[\protect\citeauthoryear{{Stonkut{\.e}}, {Vansevi{\v c}ius}, {Arimoto},
  {Hasegawa}, {Narbutis}, {Tamura}, {Jablonka}, {Ohta} \&
  {Yamada}}{{Stonkut{\.e}} et~al.}{2008}]{stonkute08}
{Stonkut{\.e}} R.,  {Vansevi{\v c}ius} V.,  {Arimoto} N.,  {Hasegawa} T.,
  {Narbutis} D.,  {Tamura} N.,  {Jablonka} P.,  {Ohta} K.,    {Yamada} Y.,
  2008, \aj, 135, 1482

\bibitem[\protect\citeauthoryear{{Tsuchiya}, {Korchagin} \&
  {Dinescu}}{{Tsuchiya} et~al.}{2004}]{tsuchiya04}
{Tsuchiya} T.,  {Korchagin} V.~I.,    {Dinescu} D.~I.,  2004, \mnras, 350, 1141

\bibitem[\protect\citeauthoryear{{van den Bergh}}{{van den
  Bergh}}{1996}]{vanden96}
{van den Bergh} S.,  1996, \aj, 112, 2634

\bibitem[\protect\citeauthoryear{{van den Bergh} \& {Mackey}}{{van den Bergh}
  \& {Mackey}}{2004}]{vanden04}
{van den Bergh} S.,  {Mackey} A.~D.,  2004, \mnras, 354, 713

\bibitem[\protect\citeauthoryear{{Walsh}, {Jerjen} \& {Willman}}{{Walsh}
  et~al.}{2007}]{walsh07}
{Walsh} S.~M.,  {Jerjen} H.,    {Willman} B.,  2007, \apjl, 662, L83

\bibitem[\protect\citeauthoryear{{Willman}, {Blanton}, {West}, {Dalcanton},
  {Hogg}, {Schneider}, {Wherry}, {Yanny} \& {Brinkmann}}{{Willman}
  et~al.}{2005}]{willman05a}
{Willman} B.,  {Blanton} M.~R.,  {West} A.~A.,  {Dalcanton} J.~J.,  {Hogg}
  D.~W.,  {Schneider} D.~P.,  {Wherry} N.,  {Yanny} B.,    {Brinkmann} J.,
  2005, \aj, 129, 2692

\bibitem[\protect\citeauthoryear{{Willman}, {Masjedi}, {Hogg}, {Dalcanton},
  {Martinez-Delgado}, {Blanton}, {West}, {Dotter} \& {Chaboyer}}{{Willman}
  et~al.}{2006}]{willman06}
{Willman} B.,  {Masjedi} M.,  {Hogg} D.~W.,  {Dalcanton} J.~J.,
  {Martinez-Delgado} D.,  {Blanton} M.,  {West} A.~A.,  {Dotter} A.,
  {Chaboyer} B.,  2006, ArXiv Astrophysics e-prints, 0603.486

\bibitem[\protect\citeauthoryear{{Zucker}}{{Zucker}}{2004}]{zucker04}
{Zucker} D.~B.~o.,  2004, \apjl, 612, L121

\end{thebibliography}
\bibliographystyle{mn2e.bst}

\begin{table*}
\begin{center}
\caption{Properties of EC4 DEIMOS fields and stars.}
\label{tableSat}
\begin{tabular}{lllccccc}
\hline
 field $\alpha$ (J2000) & field $\delta$ (J2000) & field &\# targeted stars & \# EC4 stars & {}  \\
\hline
00:58:15.50 & +38:03:01.1 & EC4 (F25/F26) & 211 & 8 & {} \\
\hline
star $\alpha$ (J2000) & star $\delta$ (J2000) & vel (kms$^{-1}$) & [Fe/H]$_{spec}$ &  [Fe/H]$_{phot}$ & S/N & $I$-mag & $V$-mag  \\
\hline
00:58:17.2 & +38:02:49.6  & -285.5$\pm$3.1    & -1.7 & -1.16 & 5.2 & 21.1 & 22.7  \\
00:58:17.1 & +38:02:54.1  & -286.8$\pm$4.1    & -1.5 & -1.36 & 5.9 & 20.9 & 22.5  \\
00:58:16.0 & +38:02:56.1  & -275.4$\pm$8.9    & -1.8 & -1.21 & 2.7 & 21.3 & 22.8  \\
00:58:15.5 & +38:02:58.9  & -285.6$\pm$14.1   & -0.5 & -1.45 & 3.0 & 20.9 & 22.4  \\
00:58:14.7 & +38:03:00.8  & -277.3$\pm$20.0   & -5.6 & -1.49 & 1.5 & 21.5 & 22.8  \\
00:58:16.0 & +38:02:22.5  & -285.8$\pm$2.3    & -1.9 & -1.41 & 8.9 & 20.4 & 22.1  \\
00:58:15.2 & +38:03:01.3  & -295.7$\pm$9.5    & -2.0 & -1.39  & 5.3 & 21.0 & 21.7  \\
00:58:14.4 & +38:03:00.4  & -294.1$\pm$3.9    & -2.2 & -1.48 & 2.9 & 21.6 & 22.9  \\
\hline
\end{tabular}
\end{center}
\end{table*}

\end{document}